\documentclass[prl,twocolumn,amsmath,amssymb,showpacs,floatfix]{revtex4}
\usepackage{graphicx}

\begin{document}




\title{Minimum Instances of Topological Matter in an Optical Plaquette}

\author{Bel\'{e}n Paredes}
\author{Immanuel Bloch}
\affiliation{Institut f\"{u}r Physik, Johannes Gutenberg-Universit\"{a}t, Staudingerweg 7, D-55099 Mainz, Germany}

\begin{abstract}
We propose experimental schemes to create and probe minimum forms of different topologically ordered states in a plaquette of an optical lattice: Resonating Valence Bond, Laughlin and string-net condensed states. We show how to create anyonic excitations on top of these liquids and detect their fractional statistics. In addition, we propose a way to design a plaquette ring-exchange interaction, the building block Hamiltonian of a lattice topological theory. Our preparation and detection schemes combine different techniques already demonstrated in experiments with atoms in optical superlattices.
\end{abstract}

\date{\today}
 \maketitle

Strong correlations between particles can lead to unconventional states of matter that break the traditional paradigms of condensed matter physics \cite{Fisher06}. 
Among these exotic phases topological liquids \cite{book_wen} are at the frontier of current theoretical and experimental research.
They are disordered states that do not break any symmetries when cooled to zero temperature. Surprisingly, they exhibit some kind of exotic order, dubbed topological order \cite{book_wen,SternAnyonsReview}, which can not be understood in terms of a local order parameter. This global hidden pattern is revealed in the peculiar behavior both of the ground state, with a degeneracy that depends on the topology of the system and of the elementary excitations, which are anyons with fractional statistics \cite{Wilczek}.

The interest in topological liquids started in connection with  two landmark phenomena in condensed matter physics: fractional quantum Hall effect \cite{book_FQHE} and high temperature superconductivity \cite{SachdevCuprates}.
In fractional quantum Hall systems electrons organize themselves in topological liquids, like the Laughlin state \cite{LaughlinState}, following a global pattern that can not be locally destroyed. 
High temperature superconductivity was proposed by Anderson \cite{Anderson87} to occur when doping a topological spin liquid: a Resonating Valence Bond (RVB) state in which the system fluctuates among many singlet bond configurations \cite{Anderson87,Kivelson87}.
Recently, the study of topological states of matter has received special attention in the context of topological quantum computation \cite{Kitaev03,SternAnyonsReview},
which seeks to exploit them to encode and manipulate information in a manner which is resistant to errors.
Moreover, understanding topological order may  help us to understand the origin of elementary particles. According to Wen's theory \cite{Levin05}, fundamental particles, like photons and electrons, may be indeed collective excitations that emerge from a topologically ordered vacuum, a string-net condensate \cite{Levin05}. 

Except for the fractional quantum Hall effect, there is no experimental evidence as to the existence of topologically ordered phases. It remains a huge challenge to develop theoretical techniques to look for topological liquids in realistic models and find them in the laboratory.
In this direction, artificial design of topological states in the versatile and highly controllable atomic systems in optical lattices \cite{ColdAtomsReview} appears to be a very promising possibility \cite{Zoller, Trebst, DasSarma06, Duan03, Santos04, Sorensen05, Buchler05}.

In this Letter, we show how to use ultracold atoms in optical lattices to create and detect different instances of topological order in the minimum non-trivial lattice system: four spins in a plaquette. Using a superlattice structure
\cite{Sebby06,Sebby07,Porto07,Simon07,Stefan07} it is possible to devise an array of disconnected plaquettes, which can be controlled and detected in parallel.
When the hopping amplitude between plaquette sites is very small, atoms are site localized and the physics is governed by the remaining spins. By combining different techniques we show how to prepare these spins in minimum versions of topical topological liquids: a RVB state,  a Laughlin state,  and a string-net condensed state. 
Making use of the experimental ability to control superexchange interactions \cite{Stefan07,Rey07} between neighboring spins a RVB state can be created \cite{Trebst}. This state can be transformed into a Lauhghlin state of two particles, which surprisingly appears in the absence of any rotation or effective magnetic field. 
This correspondence of the RVB state and the Laughlin state exactly demonstrates for the case of a plaquette the equivalence of these two states proposed by Laughlin \cite{Laughlin89}.
To stabilize a string-net condensed state we develop a way to isolate a ring-exchange interaction \cite{Buchler05} involving the four spins in the plaquette.
The superlattice structure offers a set of precise tools to characterize these states. By merging plaquettes into double wells or single wells, in combination with usual interference experiments and spin selective measurements, the properties of these states can be revealed.
Even though their minimum size, these states allow us to observe 
instances of topological order. By locally addressing each spin in a plaquette, we show how to create anyons on top of these liquids and detect their fractional statistics. 

As an additional application, a two-particle paired state with $d$-wave symmetry, the fundamental component of a high temperature superconductor \cite{SachdevCuprates}, could be created by doping a RVB state \cite{Trebst}. A novel technique allows us to reveal the characteristic $d$-wave symmetry of this state.
 
The mini-topological liquids we consider here constitute fundamental building blocks of larger topologically ordered states \cite{Ehud02, Kivelson07}. Furthermore, plaquette ring exchange interactions are the basic ingredients of lattice gauge theories \cite{GaugeTheoriesReview}, theoretical models  describing topological matter. By connecting plaquettes in the appropriate manner a variety of strongly correlated many body states could be achieved \cite{Ehud02, Kivelson07}. 

{\em The system}.
We consider a system of atoms in two internal states $\sigma=\uparrow, \downarrow$, which for the case of e.g.  $^{87}$Rb atoms could correspond to the hyperfine states $|F=1,m_F=+1\rangle$ and $|F=1,m_F=-1\rangle$. The atoms are loaded into a two dimensional superlattice, which is produced by superimposing a long and a short period lattice \cite{Simon07} both in the $x-$ and $y-$direction in such a way that an array of disconnected plaquettes is created (see Fig.~1). In the following, we will restrict the discussion to the case of bosonic atoms, though similar results can be obtained in a straightforward manner for the case of fermions. 
\begin{figure}
	\begin{center}
		\includegraphics[width=0.48\textwidth]{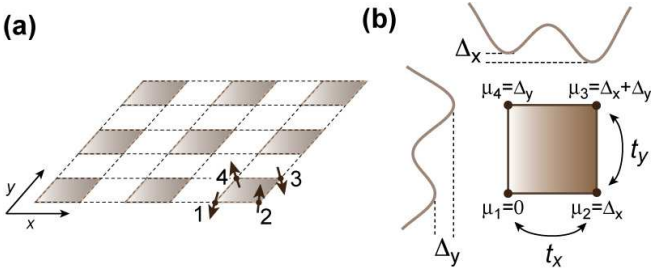}
	\end{center}
	\label{fig:PlaquetteArray}
	\caption{Schematics of optical lattice setup. Using an optical superlattice configuration along two orthogonal lattice directions, an array of decoupled plaquettes can be created {\bf (a)}. The sites within a plaquette are denumbered as in {\bf (a)}. By controlling the two optical superlattices independently, different potential bias, $\Delta_x$ and $\Delta_y$, can be introduced along the $x-$ and $y-$ direction, leading to different site energy offsets $\mu_i$ as well as different vibrational level splittings at the lattice sites.} 
\end{figure}

The dynamics of atoms in a single plaquette is governed by the Hubbard Hamiltonian
\begin{equation}
H=
-\!\!\!\! \sum_{\langle i,j\rangle,\sigma} 
 \!\!\! t_{ij} (a^{\dagger}_{i\sigma}                 
              a^{\,}_{j\sigma}+\text{H.c.})+\,
U\!\! \sum_{i,\sigma, \sigma'} 
 \!\! n_{i\sigma}n_{i\sigma'}+
 \sum_{i,\sigma}\!  \mu_{i\sigma}     n_{i\sigma},
\nonumber
\end{equation}
where $a_{i\sigma}$ and $n_{i\sigma}$ are, respectively, the bosonic annihilator and the particle number operator at site $i$ and for spin $\sigma$. 
By controlling the superlattice structure, the tunneling amplitudes in the $x$- and $y$- direction, 
$t_x \equiv t_{12}=t_{34}$ and 
$t_y \equiv t_{23}=t_{14}$, can be tuned independently. Furthermore, the dependence of the offset energies $\mu_{i\sigma}$ on position and spin state can be designed using additional magnetic offsets or gradient fields.
In the following we will make full use of the experimental ability to control these parameters, as already demonstrated in \cite{Simon07, Stefan07} for a single double well.

{\em RVB in a plaquette}.
RVB states, in which particles are paired into short-range singlets, are one of the most relevant examples of topological spin liquids \cite{SachdevCuprates}. We consider here minimum forms of RVB states consisting of four particles in a plaquette:
\begin{equation}
|\Phi_{\pm}\rangle \propto
\left(
s^{\dagger}_{1,2}s^{\dagger}_{4,3} \pm
s^{\dagger}_{1,4}s^{\dagger}_{2,3}
\right) 
|0 \rangle.
\label{RVB}
\end{equation}
Here,
$s^{\dagger}_{i,j}=
(a^{\dagger}_{i\uparrow} a^{\dagger}_{j \downarrow}
-a^{\dagger}_{i\uparrow} a^{\dagger}_{j \downarrow})$ 
creates a singlet state on sites $i$ and $j$ and $|0 \rangle$ is the vacuum state.
The states (\ref{RVB}) are disordered states with zero local magnetization, $\langle S_i^{z} \rangle=0$, for all sites $i$. They are both total singlets, with 
$S^{-}|\Phi_{\pm}\rangle=\sum_i S_i^{-}|\Phi_{\pm}\rangle=0$.
But they behave differently under rotation of the plaquette by $90^\circ$: 
$|\Phi_+\rangle$ is even (has $s$-wave symmetry), whereas $|\Phi_-\rangle$ is odd (has $d$-wave symmetry).
As larger RVB states \cite{Kivelson87, Thouless87}, states (\ref{RVB}) exhibit topological order. 

In the following we develop a scheme to prepare and detect the state $|\Phi_+\rangle$. The state $|\Phi_-\rangle$ can be  designed in a similar manner. We start with a situation in which we have four particles per plaquette and tunneling is only allowed along the $y$- direction. The system can be then prepared in a valence bond state
$|\text{VB}_y\rangle=
s^{\dagger}_{1,4}s^{\dagger}_{2,3}
|0\rangle$, 
with singlets in the vertical bonds (see Fig.~\ref{fig:StatePreparation}). 
By adiabatically turning on tunneling along the $x$- direction we will connect the state $|\text{VB}_y\rangle$ to the state $|\Phi_+\rangle$. To make this connection possible, we consider a situation in which the tunneling amplitudes $t_x$ and $t_y$ are very small in comparison to the on-site interaction energy $U$. Under these conditions, the particles are site localized and the physics is described by the superexchange interactions between the remaining spins:
\begin{figure}[b]
	\begin{center}
		\includegraphics[width=0.48\textwidth]{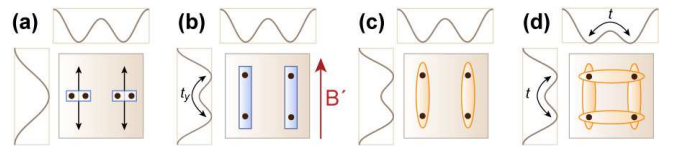}
	\end{center}
	\caption{Preparation of a RVB state. Two spin-triplet atom pairs are split into double wells along the $y$-direction in a double well-potential {\bf (a)}. A magnetic field gradient along the $y$-direction is kept for a finite amount of time to turn the spin triplet bonds along the vertical direction into spin-singlet states {\bf (b-c)}. Finally, the tunnel coupling along the $x$-direction is increased adiabatically to connect the bonds, such that a RVB state is formed {\bf (d)}. The symbols for the bonds are explained in Fig.~\ref{fig:PlaquetteEigenstatesEvolution}.} 
	\label{fig:StatePreparation}
\end{figure}
\begin{equation}
H_S=
J_x
\left(
\hat{\mathcal{P}}_{1,2}+
\hat{\mathcal{P}}_{3,4}
\right)+
J_y
\left(
\hat{\mathcal{P}}_{2,3}+
\hat{\mathcal{P}}_{1,4}
\right)+\ldots
\label{ham_spin}
\end{equation}
Here, 
$\hat{\mathcal{P}}_{i,j}=
s^{\dagger}_{i,j}
s^{\,}_{i,j}$
is the projector onto a singlet state on sites $i$ and $j$,
$J_{x(y)}=4t_{x(y)}^2/U$, and the dots denote higher order terms in $J_{x(y)}$.
If $t_x$ is suddenly increased from $0$ to $t_y$ the system will resonate with frequency $2J_y/\hbar$ between the two valence bond states 
$|\text{VB}_y\rangle$ and 
$|\text{VB}_x\rangle=s^{\dagger}_{1,2}s^{\dagger}_{4,3}|0\rangle$.
These oscillations could be detected by monitoring singlet pairs in the $x$- and $y$-bonds using a bandmapping technique \cite{Sebby07,Simon07} after merging the double wells in either the $x$- or $y$-direction into a single well. For the case of bosons, an initial spin singlet state with an antisymmetric spin wavefunction will lead to half of the population in the first excited state, whereas for a spin triplet state only the lowest vibrational state will be occupied after merging, thus allowing one to distinguish between the two spin states \cite{Rey07}. 
In order to create the state $|\Phi_+\rangle$, $t_x$ has to be turned on adiabatically. As shown in Fig.~\ref{fig:PlaquetteEigenstatesEvolution}. the states $|\text{VB}_y\rangle$ and $|\Phi_+\rangle$ are adiabatically connected. The only state to which a transition is not forbidden by symmetry constraints is  $|\Phi_-\rangle$, the other total singlet. The energy gap to this state is always on the order of $\sim 2J_y$, giving a time scale of tens of ms, which can be easily fulfilled in experiments. 
\begin{figure}
	\begin{center}		\includegraphics[width=0.45\textwidth]{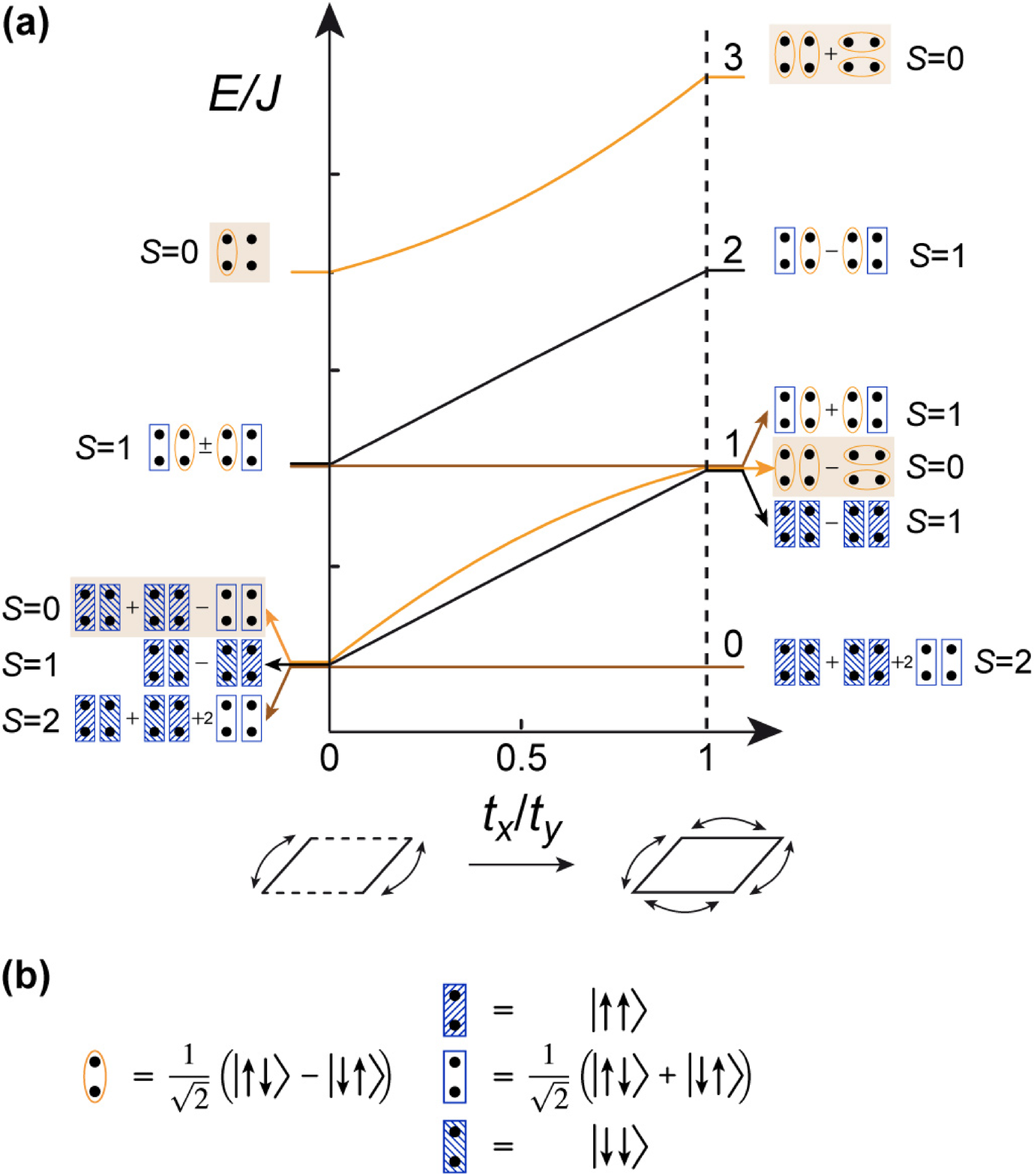}
	\end{center}
	\caption{{\bf (a)} Eigenstates and eigenenergies of the spin Hamiltonian (\ref{ham_spin}) vs. tunnel coupling ratio along the horizontal vs. vertical bond. {\bf (b)} Legend explaining the different symbols used to characterize spin singlet and spin triplet states of two particles.} \label{fig:PlaquetteEigenstatesEvolution}
\end{figure}

The two RVB states presented in eq.~(\ref{RVB}), can be characterized and differentiated from each other using two alternative methods.

{\em a) Merging technique}. By merging wells along the $x$-, $y$- or diagonal direction we can monitor singlets in that direction via the bandmapping technique outlined in \cite{Simon07}. For example, by merging along the diagonals we could easily discriminate 
$|\Phi_+\rangle=
\frac{1}{\sqrt{3}}\left(
t^{+\dagger}_{1,3}t^{-\dagger}_{2,4}+
t^{-\dagger}_{1,3}t^{+\dagger}_{2,4}-
t^{\dagger}_{1,3}t^{\dagger}_{2,4}
\right)|0\rangle$, 
which is made out of triplets in the diagonal bonds, with 
$t^{+(-)\dagger}_{i,j}=a^{\dagger}_{i\uparrow(\downarrow)} a^{\dagger}_{j \uparrow (\downarrow)}$,
from 
$|\Phi_-\rangle=s^{\dagger}_{1,3}s^{\dagger}_{2,4}|0\rangle$, made out of singlets. 
Alternatively, we can distinguish these two states by merging along the vertical direction. To do this we first undo the adiabatic path we followed before (see Fig.~\ref{fig:PlaquetteEigenstatesEvolution}) by decreasing $t_x$ from $t$ to $0$. In this way $|\Phi_+\rangle$ will be connected to the state
$s^{\dagger}_{1,2}s^{\dagger}_{4,3}|0\rangle$, with singlets in the vertical bonds, whereas $|\Phi_-\rangle$ will connect to
$\frac{1}{\sqrt{3}}\left(
t^{+\dagger}_{1,2}t^{-\dagger}_{4,3}+
t^{-\dagger}_{1,2}t^{+\dagger}_{4,3}-
t^{\dagger}_{1,2}t^{\dagger}_{4,3}
\right)
|0\rangle$, with triplets in the vertical bonds.

{\em b) Conversion into a polarized two-particle state}.
Any four spin state $|\Phi\rangle$ with well defined $S^z=\sum_i S^z_i=0$ can be written as a state of two up particles in a background of spin down particles
\begin{figure}
	\begin{center}
		\includegraphics[width=0.5\textwidth]{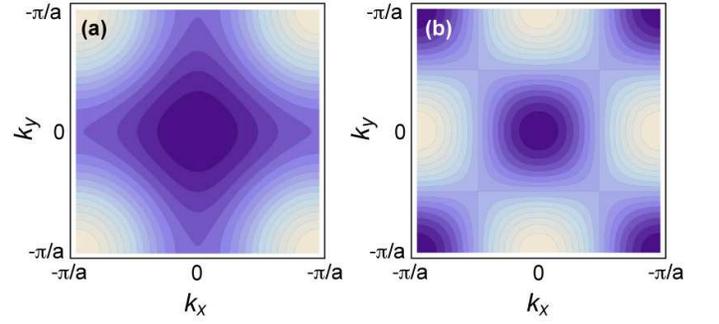}
	\end{center}

	\caption{Momentum distribution of the polarized two-particle states resulting from the RVB states $|\Phi_+\rangle$ {\bf (a)} and $|\Phi_-\rangle$ {\bf (b)} (see text). Such momentum distributions could be observed after releasing the atoms from the optical lattice potentials and a subsequent time-of-flight period.}	\label{fig:TOFMomentum} 
\end{figure}
\begin{equation}
|\Phi\rangle=
\sum_{x_1,x_2}
\psi(x_1,x_2)
S^{+}_{x_1}S^{+}_{x_2}
|\!\!\downarrow\downarrow\downarrow\downarrow \rangle,
\label{mapping}
\end{equation}
where
$S^{+}_{x}$ is the spin raising operator on site $x=1,\ldots,4$,  and
$|\!\!\downarrow\downarrow\downarrow\downarrow\rangle=
a^{\dagger}_{1\downarrow}a^{\dagger}_{2\downarrow}
a^{\dagger}_{3\downarrow}a^{\dagger}_{4\downarrow}
|0\rangle$.
If we remove the background of spin down particles, that is, if we 
apply the operator 
$\sum_{i \ne j} a_{i\downarrow}a_{j\downarrow}$ to the state (\ref{mapping}),
we are  left with a system of two polarized hard-core bosons with wave function $\psi(x_1,x_2)$.
In practice, spin down particles can be effectively removed by projecting the spin down part of state (\ref{mapping}) onto the state 
$\sum_{i \ne j}
a^{\dagger}_{i\downarrow}a^{\dagger}_{j\downarrow}|0_{\downarrow}\rangle$, 
where $|0_{\downarrow}\rangle$ is the vacuum of down particles.
The properties of the resulting two particle state directly reflect those of the spin parent state (\ref{mapping}).
By observing the momentum distribution of the two particles via a common time of flight experiment, we can read back the spin-spin correlations of the parent spin state. For example, observing a hole at the center of the time of flight picture would be an unambiguous signature of a total spin singlet parent state. To see this, note that a spin state $|\Phi\rangle$ is a total singlet if and only if $\langle\Phi| S^+S^-|\Phi\rangle=0$, which for the corresponding two-particle  state translates into a zero occupation of the momentum state with $\vec{k}=(0,0)$. Moreover, for the case of a total singlet, the spin structure factor, 
$n_{\vec{k}}
\propto
\sum_{\vec{x},\vec{x}'}
e^{i\vec{k}\cdot(\vec{x}-\vec{x}')}\langle 
\vec{S}_{\vec{x}} 
\vec{S}_{\vec{x}'}\rangle$,
(with 
$\vec{k}=\frac{\pi}{a}(\ell_x,\ell_y)$, $\vec{x}=(\ell_x,\ell_y)a$, 
$\ell_x,\ell_y=0,1$),
is directly equivalent to the momentum distribution of the corresponding two-particle state, since for a rotationally invariant state we have 
$\langle 
\vec{S}_{\vec{x}}\vec{S}_{\vec{x}'}\rangle
\propto \text{Re}
\langle S^+_{\vec{x}}S^-_{\vec{x}'}\rangle$.
Additionally, the symmetry of the spin state ($s$-wave, $p$-wave, or $d$-wave) is inherited by the two particle wave function $\psi$, and is directly reflected in the center of mass angular momentum $(L_{\text{cm}}=0,\pm 1, 2)$ of the pair.
To see this, note that the center of mass angular momentum operator, 
$\hat{L}_{\text{cm}}\propto\sum_{x=1}^4 a^{\dagger}_{x+1} a_x$,
is equivalent to the translation operator by one site along the ring, which rotates the plaquette by $90^\circ$.

This conversion technique provides us with a way to fully characterize the states (\ref{RVB}). The corresponding time of flight pictures (see Fig.~\ref{fig:TOFMomentum}) show both a dip at the center, probing their being total singlets, but are quite different in structure, a  consequence of the different symmetry of the $s$-wave and $d$-wave RVB states.
The two particle states obtained from the RVB states (\ref{RVB}) turn out to be of special importance by themselves. We show below that the state $|\Phi_+\rangle$ can be converted into a Laughlin state, whereas $|\Phi_-\rangle$ can be transformed into a paired state with $d$-wave symmetry.

\begin{figure}
	\begin{center}
		\includegraphics[width=0.4\textwidth]{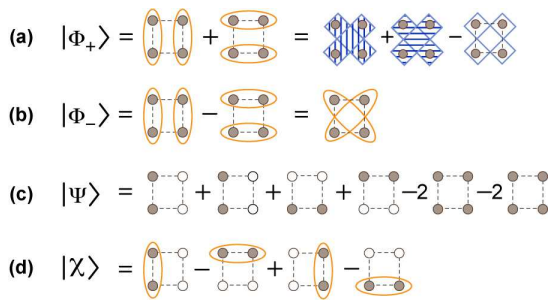}
	\end{center}
	\label{fig:States}
	\caption{Legend of correlated quantum states on a plaquette that are created using the methods described in the text. {\bf (a)} s-wave RVB state, $|\Phi_+\rangle$, {\bf (b)} d-wave RVB state, $|\Phi_-\rangle$, {\bf (c)} Laughlin state, {\bf (d)} paired state with $d$-wave symmetry.} 
\end{figure}
{\em Laughlin State}.
The Laughlin state \cite{LaughlinState} is one of the best known examples of a topologically ordered state.
Here, we describe a way to prepare a Laughlin state of two particles starting with the state $|\Phi_+\rangle$ in (\ref{RVB}). In contrast to previous schemes our method does not require either rotation \cite{Paredes01} of the optical lattice or of the individual wells \cite{Popp04}, or the presence of effective magnetic fields \cite{Sorensen05}.
The key point is to realize that the state $|\Phi_+\rangle$ can be written in the form (\ref{mapping}) with 
\begin{equation}
\psi(x_1,x_2)=z_1z_2(z_1-z_2)^2,
\label{Lau}
\end{equation}
where $z_i=e^{i\frac{\pi}{2}x_i}$, $x_i=1,\ldots,4$. 
By  removing the spin down particles, a Laughlin state of the remaining spin up particles is created (\ref{Lau}). 
This state is an eigenstate of the total angular momentum operator, $\hat{L}=\sum_m m a^{\dagger}_ma_m$, with eigenvalue $L=4$. Here, the operator
$a^{\dagger}_m=\frac{1}{2}
\sum_{\ell=1}^4e^{i\pi/4m\ell}a^{\dagger}_\ell$, creates a particle in a state of angular momentum $m$.
It has also a well defined center of mass angular momentum, $L_{\text{cm}}=0$. This vanishing $L_{\text{cm}}$ is a consequence of the s-wave symmetry of the wave function, inherited from the  state $|\Phi_+\rangle$.
The state (\ref{Lau}) is indeed a Laughlin quasihole state \cite{LaughlinState}. It contains a quasihole at the center of the plaquette, whose characteristic density profile, with a dip at the center, could be observed in a time of flight interference experiment (see Fig.~\ref{fig:PlaquetteEigenstatesEvolution}).

The equivalence of a long-range RVB state of $2N$ spins and a Laughlin state of $N$ hard-core bosons has been proposed by Laughlin for a triangular two dimensional lattice Hamiltonian \cite{Laughlin89}. Indeed, it is known that this connection is exact for a lattice of spins sitting on a ring and interacting with a long-range interaction, the so called Haldane-Shastry model \cite{HaldaneShastryModel}: 
$H_{\text{HS}}=\sum_{i,j} J_{ij}
\hat{\mathcal{P}}_{i,j}$, 
with $J_{ij}^{-2}= \sin\left[\frac{\pi}{2N}(x_i-x_j)\right]$. 
A surprising fact in our case is that the Laughlin state (\ref{Lau}) appears in the absence of any frustration or long-range interaction (there is no interaction between spins along the diagonals in Hamiltonian (\ref{ham_spin})).
The key point to understand why the Laughlin state appears nevertheless, is to realize that for the special case of a plaquette the Hamiltonian (\ref{ham_spin}) can be written in the form $H_{\text{S}}=\frac{1}{2}S^2-H_{\text{HS}}$, with $S$ being the total spin operator. Since $H_{\text{HS}}$ is rotationally invariant, we have that $[H_{\text{HS}},S^2]=0$ and  $H_{\text{S}}$ and $H_{\text{HS}}$ share the same eigenstates with well defined $S^2$.
This has important consequences for the elementary excitations of Hamiltonian (\ref{ham_spin}). As for $H_{\text{HS}}$, they are $\frac{1}{2}$-quasiholes spanned by wave functions of the form $\psi_{\eta}=(z_1-\eta)(z_2-\eta)\psi$, describing half of a boson missing at position $\eta$. These quasiholes are $\frac{1}{2}$-anyons according to the generalized definition of fractional statistics introduced by Haldane \cite{HaldaneExclusion}. In terms of spins, they are $\frac{1}{2}$-spin excitations, the so called {\em spinons} \cite{HaldaneSpinonGas, Laughlin89}.

\begin{figure}
	\begin{center}
		\includegraphics[width=0.4\textwidth]{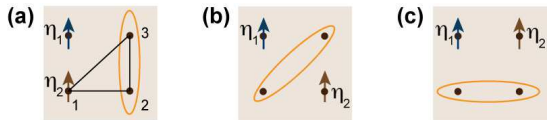}
	\end{center}
	\caption{Spinon Excitations. Linearly dependent states containing two spinon excitations (marked by the two up arrows) on a plaquette.} 
	\label{fig:Spinons}
\end{figure}

{\em Spinons and fractional statistics}.
Low-energy excitations of RVB states are created by breaking one of the spin-singlet bonds.  For the case of a plaquette, they are $S^z=1$ excitations
of the form 
$|\Phi_{\eta_1,\eta_2}\rangle=
a^{\dagger}_{\eta_1 \uparrow}a^{\dagger}_{\eta_2 \uparrow} s^{\dagger}_{x_1,x_2}|0\rangle$,
containing a pair of spinons localized at sites $\eta_1$ and $\eta_2$.
If the position $\eta_1$ of one of the spinons is fixed, there are three possible states of this form, corresponding to three different positions of the other spinon (Fig.~\ref{fig:Spinons}). It is interesting to observe that these states are not linearly independent. They generate a subspace of dimension two which is orthogonal to the state 
$\sum_{i=\eta_2,x_1,x_2}S^{-}_{i}
|\!\!\uparrow\uparrow\uparrow\uparrow \rangle$. This non-orthogonality of states describing spinons at different positions is a characteristic feature of quasiparticles with fractional statistics, as defined by Haldane \cite{HaldaneExclusion}. In contrast to the case of spinons, states describing bosons or fermions at different positions would be linearly independent.

The fractional character of spinons becomes more transparent by mapping the spin system into a hard-core boson problem, in the same way that we did above.
We consider the triangle obtained by excluding the site $\eta_1$ in which one of the spinons is fixed. By removing down particles in this triangle, the spinon state is 
mapped onto a two-particle state of the form:
\begin{equation}
\psi_{\eta_2}\propto z_1z_2
\left(\frac{\partial}{\partial z_1}-\bar{\eta}_2\right)
\left(\frac{\partial}{\partial z_2}-\bar{\eta}_2\right)
(z_1-z_2)^2,
\label{LaughlinQuasiParticle}
\end{equation} 
where $z_i=e^{i\frac{2\pi}{3}x_i}$, and $x_i=1,2,3$ enumerates the sites of the triangle in consecutive order. The state (\ref{LaughlinQuasiParticle}) describes a fractional $\frac{1}{2}$-Laughlin quasiparticle \cite{LaughlinState} located at position $\eta_2$. Since the addition of a complete boson is equivalent to a spin flip and creates an excitation with $S^z=1$, the quasiparticle, which constitutes half a boson corresponds to a spinon, with $S=1/2$.

Let us design an experiment to create spinons and probe their fractional statistics by detecting the non-orthogonality of the states $|\Phi_{\eta_1,\eta_2}\rangle$.
A state of this form  can be prepared experimentally by starting with the valence bond state
$s^{\dagger}_{\eta_1 \eta_2} s^{\dagger}_{x_1,x_2}|0\rangle$, created as explained above. If the spin of the particle at site $\eta_1$ is flipped this state is transformed into 
$\left(a^{\dagger}_{\eta_1 \uparrow}a^{\dagger}_{\eta_2 \uparrow}+
a^{\dagger}_{\eta_1 \downarrow}a^{\dagger}_{\eta_2 \downarrow}\right) s^{\dagger}_{x_1,x_2}|0\rangle$, which by measuring the spin at $\eta_1$ in the $z$ basis can be finally transformed into $|\Phi_{\eta_1,\eta_2}\rangle$.

To probe the linear dependence of these three states, we project the down part of each of them onto the state $\sum_i a^{\dagger}_{i \downarrow} |0\rangle$, a projection that will yield zero in all cases only for linearly dependent states.

{\em Paired state with $d$-wave symmetry}.
Cuprate superconductors are known to exhibit pairing with $d$-wave symmetry \cite{SachdevCuprates}. A single pair with this exotic symmetry is described by the state:
\begin{equation}
|\chi\rangle=
\frac{1}{2}
\left(
s^{\dagger}_{1,2}-
s^{\dagger}_{2,3}+
s^{\dagger}_{3,4}-
s^{\dagger}_{1,4}
\right)
|0\rangle.
\label{dwave}
\end{equation}
Let us design a scheme to create and detect this state. Starting with the four-particle state $|\Phi_-\rangle$, we first remove particles with spin down to obtain the  state
$\frac{1}{2}
\left(
t^{+\dagger}_{1,2}-
t^{+\dagger}_{2,3}+
t^{+\dagger}_{3,4}-
t^{+\dagger}_{1,4}
\right)
|0\rangle$,
a triplet pair with the desired $d$-wave symmetry. 
It is curious to see that the wave function describing this state, $\propto \bar{z}_1\bar{z}_2 (z_1+z_2)^2 (z_1-z_2)^2$, corresponds to an excited Laughlin state, with a quasiparticle excitation in the center of the plaquette (the factor $\bar{z}_1\bar{z}_2$) and an excitation of the center of mass of two units of angular momentum (the factor $(z_1+z_2)^2$). This $L_{\text{cm}}=2$ directly reflects the $d$-wave symmetry of the state. 
In order to achieve the state (\ref{dwave}) the triplet pair has to be transformed into a singlet. This can be done by using the experimental techniques demonstrated in \cite{Widera05}.
In order to reveal the $d$-wave character of the state (\ref{dwave}) we propose a novel technique which exploits the connection between the symmetry of the state and the center of mass angular momentum of the pair. By inverting the process above we transform the state (\ref{dwave}) into a spin polarized pair. 
We then merge the four sites of the plaquette into a single well and convert the pair into a molecule using a photoassociation technique \cite{Rom04}. Since angular momentum of the center of mass is conserved in the merging process, the molecule will carry two units of angular momentum, which will directly reflect the $d$-wave symmetry of the state (\ref{dwave}). 

The paired state with $d$-wave symmetry (\ref{dwave}) can be converted into a pair with a non-vanishing $d$-wave order parameter through e.g. admixture of a vacuum state. This could be done, by adiabtically increasing the tunnelling to an unoccupied layer of empty plaquettes, below or above the occupied plaquette layer.

{\em Ring-exchange interactions}.
\begin{figure}
	\begin{center}
		\includegraphics[width=0.45\textwidth]{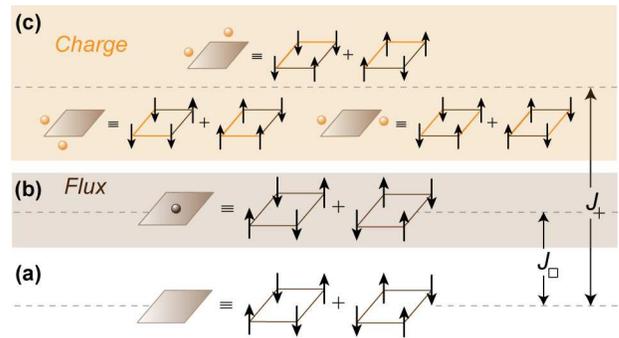}
	\end{center}
	\caption{Eigenstates of the gauge lattice Hamiltonian (\ref{hring}) (see text). Ground state {\bf(a)}. Flux quasiparticle excitation {\bf(b)}. Charge quasiparticle excitation, in which two bonds are excited {\bf(c)}.\label{fig:FluxChargeStates}
	} 
\end{figure}
Lattice gauge theories \cite{GaugeTheoriesReview} play an essential role in describing topological matter \cite{SternAnyonsReview}. 
The minimum lattice gauge Hamiltonian describes a system of four spins in a plaquette and has the form:
\begin{equation}
H_G=-J_{\square}\,\,S^{x}_1S^{x}_2S^{x}_3S^{x}_4
+J_+\sum_{<i,j>}S^{z}_iS^{z}_j.
\label{hring}
\end{equation}
It consists of four terms that commute with each other. The first one is a ring-exchange or {\em flux} interaction involving the four spins. For $J_{\square}<0$ it favors symmetric states with respect to spin flipping of the whole plaquette. The other ones are {\em charge} interactions between neighboring spins, which for $J_+>0$ favor states with anti-parallel 
neighboring spins. 
As for the case of an infinite lattice \cite{Kitaev03} the elementary excitations of this Hamiltonian are anyons. Though this is a well known result, for the sake of clarity of our discussion bellow, let us first briefly explain it for the case of a single plaquette.
The ground state of Hamiltonian (\ref{hring}) is a GHZ state of the form
\begin{equation}
|\square  \rangle= \frac{1}{\sqrt{2}} \left(
|\!\! \uparrow \downarrow \uparrow \downarrow \rangle
+|\!\! \downarrow \uparrow \downarrow \uparrow \rangle
\right),
\label{GHZ}
\end{equation} 
a maximally entangled state of four particles. It is indeed the minimum version of a string-net condensate \cite{Levin05}, the ground state of (\ref{hring}) when extended to an infinite lattice.
We can create two types of excitations on top of the state (\ref{GHZ}). They are flux-like or charge-like quasiparticles, (see Fig.~\ref{fig:FluxChargeStates})  depending on which term of the Hamiltonian (\ref{hring}) is excited.
For example, a flux-like excitation (fluxon), which we denote by $|\boxdot \rangle$, has the form
$|\boxdot \rangle=
\frac{1}{\sqrt{2}} \left(
|\!\! \uparrow \downarrow \uparrow \downarrow \rangle
-|\!\!\downarrow \uparrow \downarrow \uparrow \rangle
\right)$. 
It can be obtained by applying, for example, the operator 
$S^{z}_1$ to the state $|\square  \rangle$.
Charge-like excitations, in which two neighboring spins become parallel, are always created in pairs. For example, the state 
$| \!\! \cdot \!\!\dot {\square} \rangle=
\frac{1}{\sqrt{2}} \left(
|\!\!\uparrow \downarrow \uparrow \uparrow\rangle 
+|\!\!\downarrow \uparrow \downarrow \downarrow \rangle
\right)$ 
contains two charge-like quasiparticles, one at the 1-4 bond and the other at the 3-4 bond. This state is obtained by applying the operator 
$S^{x}_4$ to the state $|\square  \rangle$. 
A charge-like quasiparticle can be moved around a flux-like one 
(see Fig.~\ref{fig:Anyons}) by applying the ring operator $S^{x}_1S^{x}_2S^{x}_3S^{x}_4$ onto the state $|\boxdot \rangle$. Since
$S^{x}_1S^{x}_2S^{x}_3S^{x}_4
|\boxdot \rangle=
S^{x}_1S^{x}_2S^{x}_3S^{x}_4S^{z}_1
|\square \rangle
=-|\boxdot \rangle$, the wave function picks up a minus sign during the process.
Therefore charges and fluxons are relative $\frac{1}{2}$-anyons in this model.

\begin{figure}
	\begin{center}
		\includegraphics[width=0.45\textwidth]{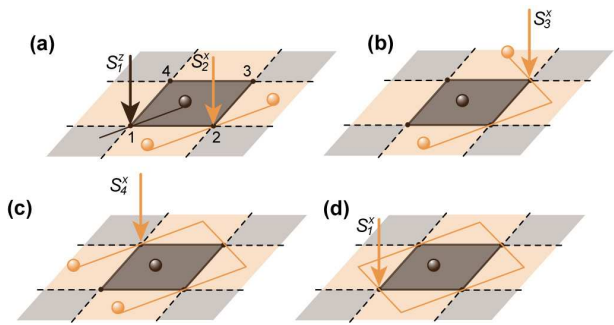}
	\end{center}
	\caption{Anyon Braiding. A flux-type and a charge-type quasiparticle are created  by applying the operators $S_1^{z}$ and $S_2^{x}$, respectively, to the ground state ({\bf a}). A single charge-type quasiparticle is then moved around the flux quasiparticle by subsequent application of the local operators $S_3^{x}$  ({\bf b}), $S_4^{x}$ ({\bf c}), and $S_1^{x}$ ({\bf d}).\label{fig:Anyons}} 
\end{figure}

In our optical plaquette a Hamiltonian like (\ref{hring}) seems, in principle, is difficult to implement. The reason behind is that four-spin interactions result from fourth order processes (higher order terms denoted by dots in equation (\ref{ham_spin})) in which four tunneling events occur. These processes are usually highly suppressed ($\sim t^4/U^3$) compared to second order processes ($\sim t^2/U$), leading to dominant next neighbor superexchange interactions \cite{Stefan07, Rey07}. 
Here, we present a scheme to suppress second order processes in a plaquette, obtaining a dominating four-body interaction. This will allow us to implement Hamiltonian (\ref{hring}) within a certain subspace of the spin Hilbert space.
\begin{figure}
	\begin{center}
		\includegraphics[width=0.28\textwidth]{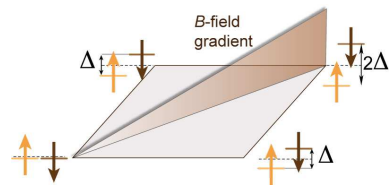}
	\end{center}
	\caption{By applying a magnetic field gradient along a diagonal direction of the plaquette, superexchange interactions can be suppressed.} 
	\label{fig:FieldGradient}
\end{figure}
\begin{figure}[b]
	\begin{center}
		\includegraphics[width=0.4\textwidth]{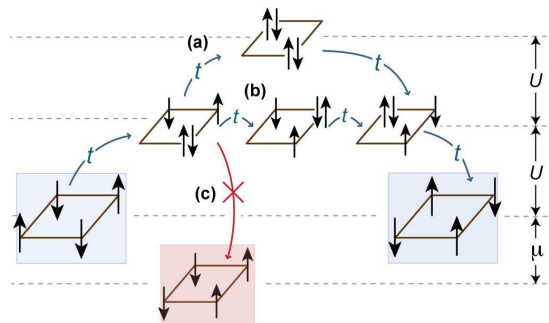}
	\end{center}
	\label{fig:RingExchange}
	\caption{Effective ring exchange interaction in the presence of a magnetic field gradient. Fourth order processes connecting the states 
$|\! \!\! \uparrow \downarrow \uparrow \downarrow \rangle$ and 
$|\! \!\! \downarrow \uparrow \downarrow \uparrow \rangle$ are resonant {\bf(a-b)}. Second order processes connecting the states 
$|\!\! \uparrow \downarrow \uparrow \downarrow \rangle$
and
$|\!\! \downarrow \uparrow \uparrow \downarrow \rangle$
are off resonance {\bf(c)}.} 
\end{figure}
We consider a situation in which we have applied a magnetic field gradient  $\Delta$ along one of the diagonals (e.g., 1-3) of the plaquette (see Fig.~\ref{fig:FieldGradient}). 
If $\Delta \gg 4t^2/U$, spin exchange interactions between neighboring sites are suppressed. The only remaining processes are either the ones in which the four spins in the plaquette are flipped (see Fig.\ref{fig:RingExchange}) or those in which spins along the diagonal 2-4 are exchanged, giving rise to the Hamiltonian:
\begin{eqnarray}
H_R&=&-J_{\square}\!\left(S^{+}_1S^{-}_2S^{+}_3S^{-}_4 +H.c.\right)
+J_+ \!\!\sum_{<i,j>}S^{z}_iS^{z}_j\nonumber\\ 
&&+J_\times \left( S^{+}_2S^{-}_4 + H.c.\right)-\Delta\sum_i B_iS^{z}_i,
\label{hreal}
\end{eqnarray}
where $J_{\square}\approx 24t^4/U^3$, $J_+\approx 4t^2/U$, $J_\times \approx 16 t^4/U^3$, and $B_1=0,B_2=B_4=1,B_3=2$.

\begin{figure}
	\begin{center}
		\includegraphics[width=0.5\textwidth]{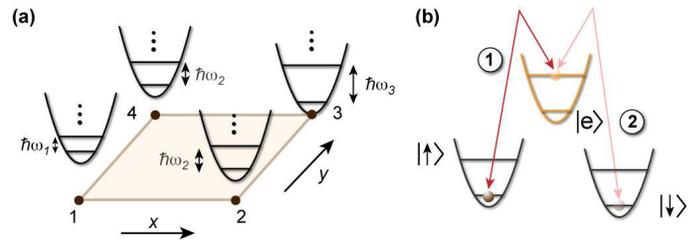}
	\end{center}
	\caption{Single site addressing within a plaquette. By adjusting the superlattice potentials in the $x$- and $y$-direction, the potential wells on the edges of a plaquette can obtain different vibrational splittings {\bf (a)}, e.g. $\hbar \omega_1 \neq \hbar \omega_2 \neq \hbar \omega_3$. This can be exploited to target the spin on a single site and manipulate it without affecting the neighboring spins in the plaquette. In order to achieve this, Raman transitions {\bf (b)} resonant to an intermediate excited vibrational state on a plaquette edge can be used. For sufficiently spectrally narrowband Raman pulses, the transitions will only be driven on a chosen single plaquette site. \label{fig:Addressing}} 
\end{figure}

Within the subspace generated by the states 
$|\!\! \uparrow \downarrow \uparrow \downarrow \rangle$ and
$|\!\! \downarrow \uparrow \downarrow \uparrow \rangle$,
Hamiltonian (\ref{hreal}) is equivalent to Hamiltonian (\ref{hring}).
The string-net condensed state $|\square \rangle$ and the flux excitation $|\boxdot \rangle$ are therefore eigenstates of our system. Let us show how to prepare this states.
We proceed as follows.
The plaquette is initially prepared in the state 
$|\!\! \uparrow \downarrow \uparrow \downarrow \rangle$.
This can be done by starting with the state 
$|\!\! \uparrow\uparrow\uparrow\uparrow\rangle$, and then spin flipping atoms on the diagonal sites by addressing them with the scheme presented in Fig.~\ref{fig:Addressing}.
In the presence of the magnetic field gradient, the system will evolve under  Hamiltonian (\ref{hring}), oscillating between the states 
$|\!\! \uparrow \downarrow \uparrow \downarrow \rangle$ and
$|\!\! \downarrow \uparrow \downarrow \uparrow \rangle$
with a frequency $\omega=J_{\square}/\hbar$. 
For typical experimental parameters this frequency is of the order of $20$\,Hz, and can be resolved experimentally \cite{Simon07, Stefan07}.
After an evolution time $T=\pi/4\omega$ the system will be prepared in the state
$\frac{1}{\sqrt{2}} \left(
|\!\! \uparrow \downarrow \uparrow \downarrow \rangle
+i |\!\! \downarrow \uparrow \downarrow \uparrow \rangle
\right)$,
a maximally entangled state that can be easily transformed into either
$|\square \rangle$ or $|\boxdot \rangle$.
This can be done by applying the local phase operator
$R_\theta=e^{i\theta(S^z_1+S^z_3)}$, with $\theta=\pi/4(-\pi/4)$,
which is performed by addressing sites $1$ and $3$, and letting the system evolve in the presence of a magnetic field $B$ in the $z$ direction for a time $T=\theta \hbar/B$. 
We can use Hamiltonian (\ref{hreal}) together with local addressability of the plaquette sites to artificially create and detect the anyonic quasiparticles of Hamiltonian (\ref{hring}). 
Our proposal has the same spirit of the one recently proposed in \cite{Pachos}, where anyonic states are artificially encoded using four photons. 
Even though anyonic states are as in \cite{Pachos} not eigenstates of our system the preparation and detection scheme we present here can be used in cases in which the Hamiltonian (\ref{hring}) may be achieved using other methods 
\cite{Buchler05}.
Our scheme follows the idea proposed in \cite{Paredes01} for anyon detection in small rotating atomic gases.
{\em a) Initialization}.
We prepare the system in the state 
$\frac{1}{\sqrt{2}}\left(|\square\rangle-i|\boxdot\rangle\right)$, a superposition of a non-excited and a flux-excited plaquette. 
Such superposition state results indeed from time evolution of the state $|\!\!\uparrow \downarrow \uparrow \downarrow \rangle$ under Hamiltonian (\ref{hreal}) after a time $T=\pi/4\omega$.
{\em b) Statistical phase accumulation.}
We then excite a pair of charge-like excitations and move one of them around the plaquette. This operation is performed by the operator $S^{x}_1S^{x}_2S^{x}_3S^{x}_4$ (see Fig.~\ref{fig:Anyons}), which we apply by subsequently addressing and acting on each site of the plaquette.
Because of the relative $\frac{1}{2}$-statistical phase of anyons, the state $|\boxdot\rangle$ will pick up a minus sign, and the system will end up in the state
$\frac{1}{\sqrt{2}}\left(|\square\rangle+i|\boxdot\rangle \right)$. 
{\em c) Detection}.
We finally let the system evolve under Hamiltonian (\ref{hring}) for a time $T=\pi/4\omega$, obtaining the final state $|\!\!\uparrow \downarrow \uparrow \downarrow \rangle$.
If the excitations happened to be bosons or fermions with trivial statistics, the final state would have been 
$|\!\!\uparrow \downarrow \uparrow \downarrow \rangle$.
These two states can be easily discriminated by, for example, measuring $S_1^{z}$.

In conclusion, we have presented a collection of schemes to create and detect instances of topological order in a minimum system: a plaquette filled with two or four particles in an optical lattice potential. Many of these could be directly implemented in current experiments using the presently available manipulation and detection techniques. 
Furthermore, the plaquette Hamiltonians we have considered could be used as unit operations to, together with an increased optical resolution to resolve individual plaquettes, engineer large scale topological liquids.

We would like to thank S.~F\"{o}lling, S.~Trotzky, J.~Pachos, and E.~Demler for helpful discussion. Furthermore, we would like to acknowledge financial support by the EU under IP (SCALA), the DFG and AFOSR under contract FA-8655-07-1-3090.

\end{document}